\documentclass[10pt,a4paper,twocolumn]{article}
\usepackage[latin1]{inputenc}
\usepackage{amsmath,amsfonts,amssymb,graphicx}
\usepackage{authblk}

\begin{document}

\title{\textbf{Stochastic resonance controlled upregulation of internal noise after hearing loss as a putative correlate of tinnitus-related neuronal hyperactivity}}
\author[1,2]{Patrick Krauss}
\author[1]{Konstantin Tziridis}
\author[1,2]{Achim Schilling}
\author[2]{Claus Metzner}
\author[1]{\\ Holger Schulze*}
\affil[1]{\small{Experimental Otolaryngology, ENT-Hospital, Head and Neck Surgery, Friedrich-Alexander University Erlangen (FAU), Germany}}
\affil[2]{Department of Physics, Center for Medical Physics and Technology, Biophysics Group, Friedrich-Alexander University Erlangen (FAU), Germany}
\affil[ ]{ }
\affil[ ]{\large{\textit{holger.schulze@uk-erlangen.de}}}
\renewcommand\Authands{ and }

\twocolumn[
\begin{@twocolumnfalse} 

\maketitle

\section*{Abstract}

\begin{abstract}
	\textbf{Subjective tinnitus (ST) is generally assumed to be a consequence of hearing loss (HL). In animal studies acoustic trauma can lead to behavioral signs of ST, in human studies ST patients without increased hearing thresholds were found to suffer from so called hidden HL. Additionally, ST is correlated with pathologically increased spontaneous firing rates and neuronal hyperactivity (NH) along the auditory pathway. Homeostatic plasticity (HP) has been proposed as a compensation mechanism leading to the development of NH, arguing that after HL initially decreased mean firing rates of neurons are subsequently restored by increased spontaneous rates. However all HP models fundamentally lack explanatory power since the function of keeping mean firing rate constant remains elusive as does the benefit this might have in terms of information processing. Furthermore the neural circuitry being able to perform the comparison of preferred with actual mean firing rate remains unclear. Here we propose an entirely new interpretation of ST related development of NH in terms of information theory. We suggest that stochastic resonance (SR) plays a key role in short- and long-term plasticity within the auditory system and is the ultimate cause of NH and ST. SR has been found ubiquitous in neuroscience and refers to the phenomenon that sub-threshold, unperceivable signals can be transmitted by adding noise to sensor input. We argue that after HL, SR serves to lift signals above the increased hearing threshold, hence subsequently decreasing thresholds again. The increased amount of internal noise is the correlate of the NH, which finally leads to the development of ST, due to neuronal plasticity along the auditory pathway. We demonstrate the plausibility of our hypothesis by using a computational model and provide exemplarily findings of human and animal studies that are consistent with our model.} 
\end{abstract}

\end{@twocolumnfalse}]

\section*{Introduction}

It is estimated that around 10 to 15 percent of the general population suffer from subjective tinnitus. This phantom percept can be very severe for affected patients and lead to insomnia, psychological disorders or even suicide (Coles, 1984; Lewis et al., 1994; Langguth et al., 2011). Tinnitus is often accompanied by a hearing loss (Heller, 2003) and recent animal studies indicate that even a relatively mild acoustic trauma may lead to a massive loss of synapses (synaptopathy) causing a so called hidden hearing loss (Liberman et al., 2015). However, an effective cure for tinnitus still does not exist, since the exact mechanisms within the auditory pathway leading to the development of tinnitus are still debated (Gerken, 1996; Eggermont, 2003; Eggermont and Roberts, 2004; Engineer et al., 2011; Knipper et al., 2011; Schaette and McAlpine, 2011; Ruttiger et al., 2013).

Some models propose a mechanism within the central auditory system called homeostatic plasticity (HP): Prolonged changes in the mean firing rates of the auditory nerve are assumed to cause HP, thereby restoring these mean firing rates and thus compensating for reduced cochlear input. This gain increase leads to increased spontaneous rates which are hypothesized to be the correlate of hyperactivity (Schaette and Kempter, 2006; Knipper et al., 2011; Schaette and McAlpine, 2011). However all HP models fundamentally lack explanatory power since it remains totally elusive why the mean firing rate of neurons should be kept constant and which benefit this might have in terms of information processing. Furthermore a concrete neural circuitry being able to perform the comparison of the preferred with the actual mean firing rate remains unclear.

Here we propose an entirely new information theory inspired interpretation of tinnitus-related development of neuronal hyperactivity. Stochastic resonance (SR) refers to the phenomenon that sub-threshold unperceivable signals can be transmitted by adding noise to the sensor input (Benzi et al., 1981; Collins et al., 1996; Levin et al., 1996; Gammaitoni et al., 1998). SR has been found ubiquitous in nature covering a wide range of systems in physical, technological and biological contexts (Wiesenfeld and Moss, 1995) and especially within the context of neuroscience (Faisal et al., 2008; Mino et al., 2014; Douglass et al., 1993). In addition, the existence of an optimal, non-zero intensity for the added noise has been demonstrated, allowing maximization of information transmission (Wiesenfeld and Moss, 1995). In self-adaptive signal detection systems based on SR, the optimum noise level is continuously adjusted via a feed-back loop, so that the system response in terms of information throughput remains optimal, even if the properties of the input signal change. For this processing principle the term adaptive SR has been coined (Mitaim et al., 1998; Mitaim et al., 2004; Wenning et al., 2003). An objective function to quantify information content is the mutual information (MI) between the sensor input and output (Shannon, 1948). In the context of SR the MI is frequently used in theoretical approaches (Levin et al., 1996; Moss et al., 2004; Mitaim et al., 2004). The choice of the MI is natural since the fundamental purpose of any transducer is to transmit information into a subsequent information processing system. It has been shown previously that the MI as a function of noise intensity has a well-defined peak that indicates the ideal level of noise to be added to the input signal (Moss et al., 2004). However, a fundamental drawback of the MI is the impossibility of calculating it in any application of adaptive SR where the signal to be detected is unknown (Krauss et al., 2015). Furthermore, even if the underlying signal is known, the use of the MI still seems to be rather impractical within the context of neural network architectures, since calculating the MI requires evaluating probability distributions, logarithms, products and fractions. In a previous work we were able to show that this fundamental drawback can be overcome by another objective function, namely the autocorrelation (AC) of the detector response. Maximizing the output AC leads to similar or even identical estimates of optimal noise intensities for SR as the MI, yet with the decisive advantage that no knowledge of the input signal is required (Krauss et al., 2015). In addition, the evaluation of AC functions may be quite easily implemented within neural networks using delay-lines and coincidence detectors (Licklider, 1951). Remarkably, a neural architecture, resembling the aforementioned delay-lines is found in the dorsal cochlear nucleus (DCN) (Osen, 1988; Hackney et al., 1990). In addition, the DCN is the earliest processing stage in the auditory pathway where tinnitus-related changes have been observed. Acoustic trauma leads to increased spontaneous firing rates in the DCN (Kaltenbach et al., 1998, 2000; Kaltenbach and Afrman, 2000; Brozoski et al., 2002; Kaltenbach et al., 2004), whereby the amount of this increase is correlated to the strength of the behavioral signs for tinnitus (Kaltenbach et al., 2004) and this hyperactivity is only found in regions innervated by the damaged parts of the cochlear receptor epithelium (Kaltenbach et al., 2002).

Summing up, we propose that adaptive SR based on maximizing the AC of the cochlear output is a major processing principle in the auditory system and operates both on short and long time scales to maintain maximum information detection even though the statistics of the input (sound intensities) are changing. If, due to acoustic trauma, the cochlear output is reduced then the AC calculated within the DCN decreases. Hence the internal noise generated within the DCN increases to improve information throughput again. In the case of chronic cochlear damages the internal noise is increased permanently, which is the correlate of hyperactivity often associated with subjective tinnitus.

\section*{Methods}
We implemented a phenomenological model of the acoustic stimuli, the auditory nerve responses and the effects of cochlear damage to auditory nerve responses. Furthermore, we model the adaptive SR principle based on mean AC in terms of coarse-grained functional units within the DCN. We focus on input-output mappings and not on single neuron models or concrete neural network architectures. Nevertheless we emphasize that each part of the adaptive SR feedback-loop is highly biologically plausible and may be implemented in a more fine-grained model.

\paragraph*{Distribution of sound intensities}
According to (Schaette et al., 2006) we assume the probability density function of the sound intensity levels $I$ (in dB) to be Gaussian with a mean value of 40 dB and standard deviation of 25 dB. In contrast to the aforementioned study, as input to the model we do not draw independent samples from this distribution during simulation but instead used an Ornstein-Uhlenbeck process (Uhlenbeck and Ornstein, 1930) in order to generate a correlated time series of intensity levels, yet with identical mean value and standard deviation (figure 1).

\begin{figure}[htb]
	\centering
	\includegraphics[width=1.0\linewidth]{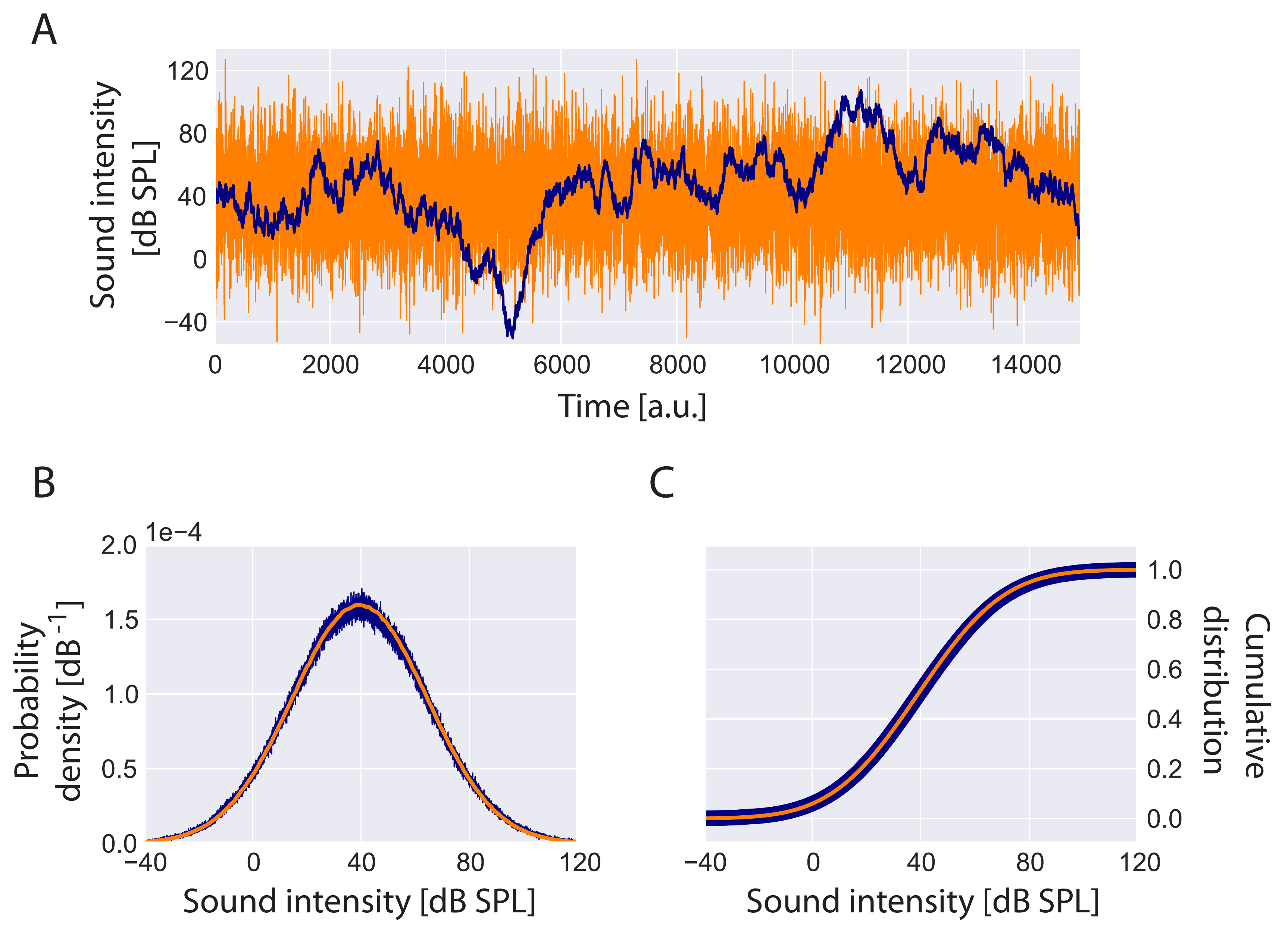}
	\caption{\em \small Sound intensities. Two sample time series of sound intensities are shown (A). The uncorrelated time series (orange) has been generated by drawing values from a Gaussian distribution, whereas the correlated time series (blue) is derived from an Ornstein-Uhlenbeck process. Although both time series look very different their probability density functions (B) and cumulative distributions (C) are identical.}
	\label{fig:1}
\end{figure}

\paragraph*{Auditory nerve responses}
The firing rate $f(I)$ of the auditory nerve at a sound intensity $I$ is modelled analogous to (Schaette et al., 2006) with a threshold $I_{th}$ of 0 dB, spontaneous firing rate $f_{sp}$ of 50 Hz and maximum firing rate $f_{max}$ of 250 Hz. The response function $f(I)$ is assumed to be adapted to the distribution of sound intensities. For $I>I_{th}$, $f(I)$ is proportional to the normalized cumulative distribution function of the sound intensities hence, according to the infomax principle, $f(I)$ has maximum information on $I$ (Laughlin, 1981). In scope of this preprint article we focus on changes to the threshold $I_{th}$ due to cochlear damage only and do not take into account changes of the spontaneous firing rate $f_{sp}$ or the maximum firing rate $f_{max}$. In figure 2 some example rate-intensity functions are shown for different thresholds.

\begin{figure}[htb]
	\centering
	\includegraphics[width=1.0\linewidth]{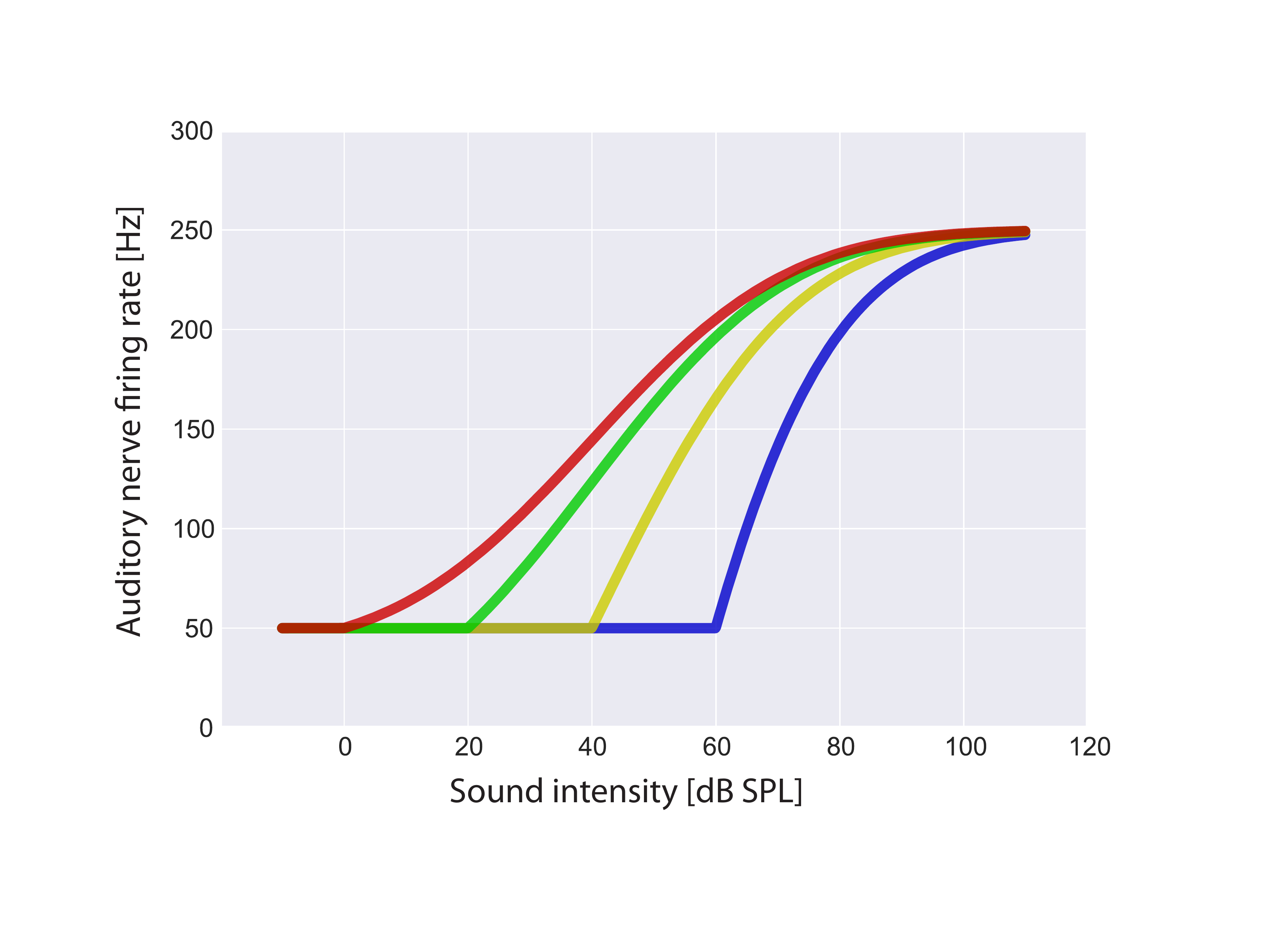}
	\caption{\em \small Rate-intensity-functions. Shown are sample rate-intensity-functions for different thresholds $I_{th}$ at 0 dB (red), 20 dB (green), 40 dB (yellow) and 60 dB (blue). In each case the spontaneous firing rate $f_{sp}$ is 50 Hz and the maximum firing rate $f_{max}$ is 250 Hz.}
	\label{fig:2}
\end{figure}

\paragraph*{Autocorrelation function and mean AC of auditory nerve firing rates}
We used the normalized autocorrelation of auditory nerve firing rates $f(t)$ as a function of the time lag $\tau$ which is defined as
\begin{equation}
C_{ff}(\tau) = \left\langle 
	(f(t)\;f(t\!+\!\tau)
	\right\rangle_t
\end{equation}
where $\left\langle \cdot \right\rangle_t$ indicates averaging over time. In order to derive a single value from this function, the mean of the autocorrelation function
\begin{equation}
\overline{C_{ff}} = \left\langle 
C_{ff}(\tau)
\right\rangle_{\tau}
\end{equation}
is calculated, where $\left\langle \cdot \right\rangle_{\tau}$ indicates averaging over all lag-times.

\paragraph*{Adaptive stochastic resonance model}
Our adaptive SR model mainly consists of two functional units that build up a feedback-loop. The first unit receives input from the cochlea and calculates the AC, reflecting the information content, of the time course of auditory nerve firing rates. We refer to this unit as the information detector (ID). The second unit is controlled by the ID and injects noise to the sensory epithelium within the cochlea via efferent connections. For this part of the system we use the term noise generator (NG) (figure 3).

\begin{figure}[htb]
	\centering
	\includegraphics[width=1.0\linewidth]{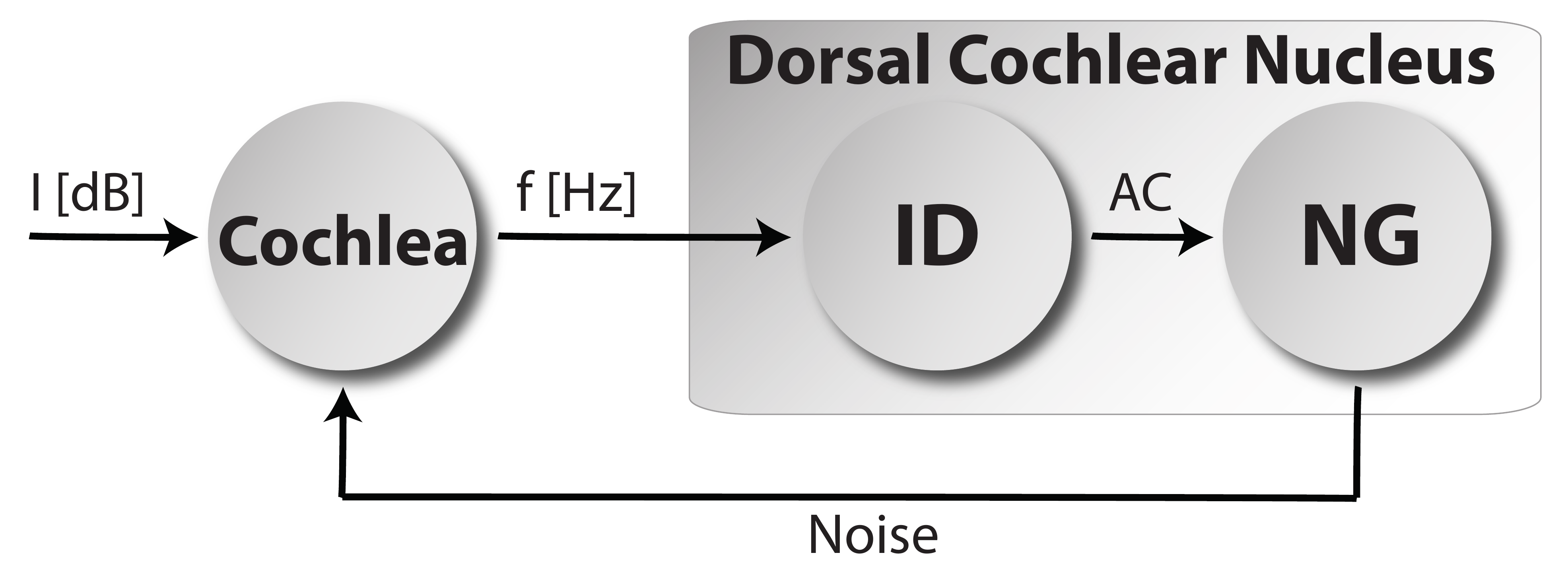}
	\caption{\em \small Adaptive SR model. The model mainly consists of two functional units that build up a feedback-loop. The information detector (ID) calculates the AC, reflecting the information content, of the time course of auditory nerve firing rates coming from the cochlea. The noise generator (NG) is controlled by the ID and injects noise to the cochlea via feedback connections. We propose that both, the ID and the NG are located within the dorsal cochlear nucleus.}
	\label{fig:3}
\end{figure}

\paragraph*{Implementation}
The model was implemented on a standard desktop PC using the programming language C/C++.

\section*{Results}
The aim of this study was to demonstrate how adaptive SR based on maximizing the AC may lead to permanently increased internal noise after chronic cochlear damages. Therefore we first present the effect of increased thresholds to the AC function of the time series of auditory nerve firing rates $f(t)$ and subsequently the benefit of SR to the AC.

\paragraph*{Damage to the cochlea decreases the mean AC}
We evaluate the AC function of the time series of auditory nerve firing rates $f(t)$ as defined in the method section. In figure 4A autocorrelation functions for different thresholds are shown. For increased thresholds the values of the AC function systematically shift to smaller values, reflecting the decreased amount of information content perceived by the cochlea. The mean AC obtained by averaging the AC function over all evaluated lag-times decreases monotonically with increasing threshold (figure 4B).

\begin{figure}[htb]
	\centering
	\includegraphics[width=1.0\linewidth]{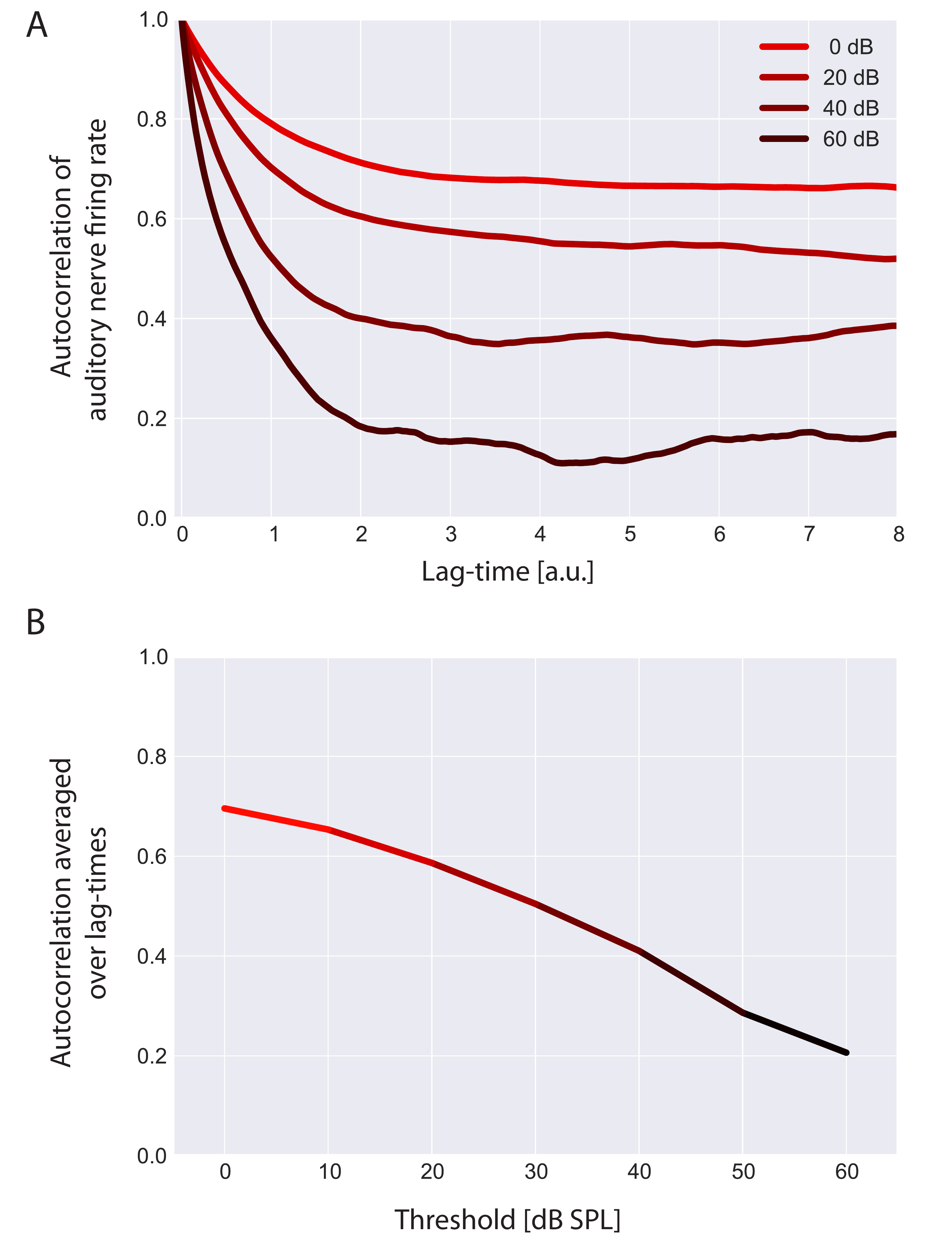}
	\caption{\em \small Autocorrelation function of the cochlear output. Shown are the AC functions for different thresholds (A). For increased thresholds the values of the AC function systematically shift to smaller values, reflecting the decreased amount of information content perceived by the cochlea. The mean AC obtained by averaging the AC function over all evaluated lag-times decreases monotonically with increasing threshold (B).}
	\label{fig:4}
\end{figure}

\paragraph*{SR improves mean AC and increases internal noise after hearing loss}
The beneficial effect of SR, as well as the increase of internal noise are shown in figure 5. In figure 5A a sample AC function for a threshold $I_{th}$ of 30 dB is shown with (green) and without (red) the effect of SR. As can be clearly seen, SR is able to significantly improve the AC. The beneficial effect of SR for different thresholds is summarized in figure 5B. With increasing thresholds more noise is required to improve the AC (figure 5C). This permanently increased noise is according to our hypothesis the correlate of tinnitus-related neuronal hyperactivity.

\begin{figure}[htb]
	\centering
	\includegraphics[width=1.0\linewidth]{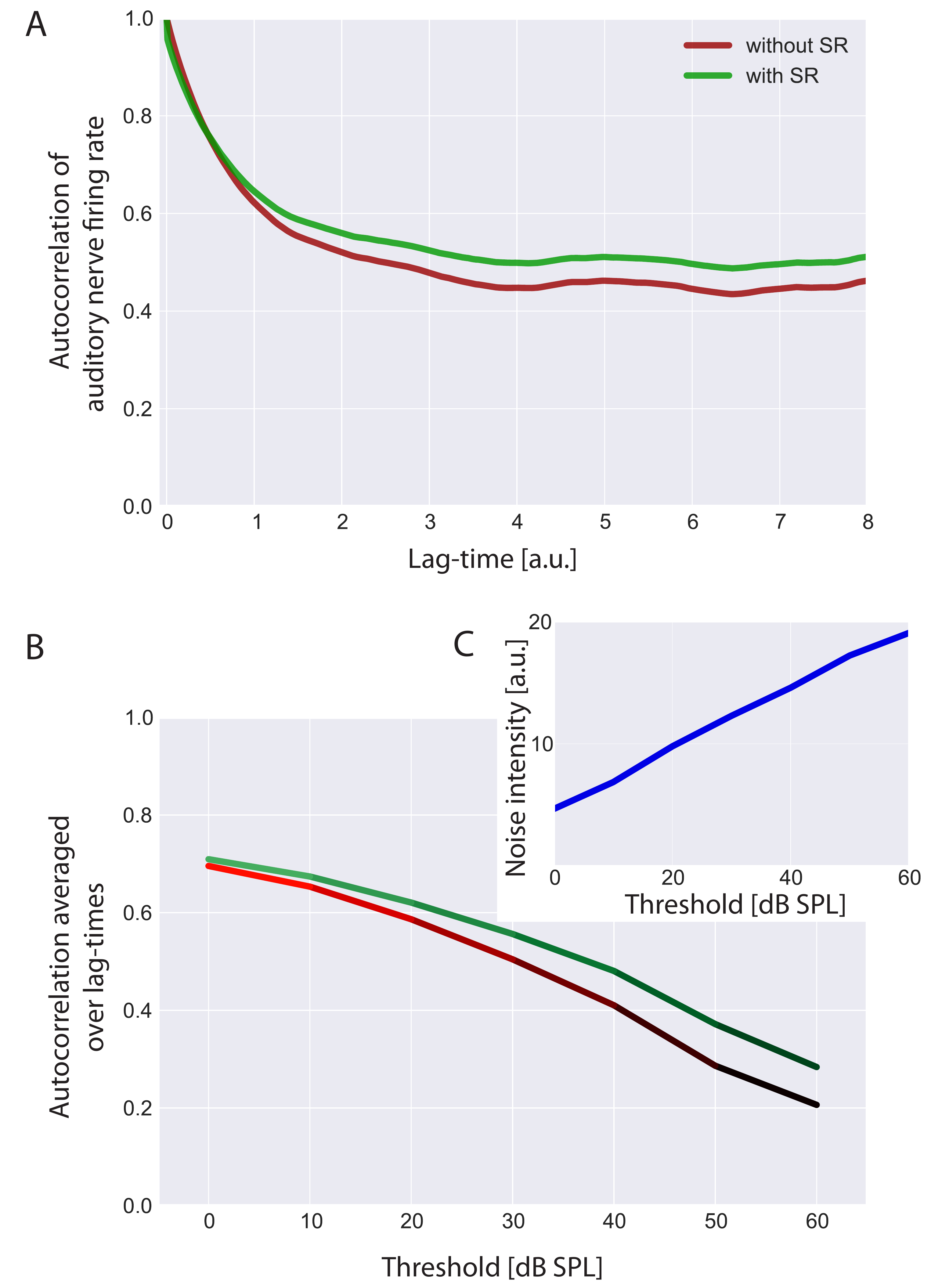}
	\caption{\em \small Effect of stochastic resonance. In (A) a sample AC function for a threshold $I_{th}$ of 30dB is shown with (green) and without (red) the effect of SR. As can be clearly seen, SR is able to significantly improve the AC. The beneficial effect of SR for different thresholds is summarized in (B). With increasing thresholds more noise is required to improve the AC (figure 5C).}
	\label{fig:5}
\end{figure}

\section*{Discussion}
We demonstrated how adaptive SR based on maximizing the AC of the cochlear output may cause neuronal hyperactivity in the case of chronic cochlear damages. This hyperactivty, if permanent, is thought to induce neuroplastic changes along the auditory pathway that subsequently lead to the development of subjective, central tinnitus. Furthermore, we suppose the DCN to be the site where SR is controlled. This is plausible since  the neuronal architecture, i.e. delay lines and coincidence detecting neurons, required to calculate the AC are actually found within the DCN. Furthermore the DCN receives massive non-auditory input from somatosensory nuclei which may serve as some kind of internal noise, which is also required for SR.

From human studies it is known that the presentation of external noise simultaneous to pure tones is actually able to significantly improve the hearing thresholds for the pure tones via SR (Zeng et al., 2000; Ries, 2007).

Another aspect that supports our hypothesis is the so called Zwicker tone illusion. The term describes an intriguing auditory aftereffect. The typical
sound generating it is a broadband noise containing a spectral gap, which is presented for several seconds. After the noise has been switched off, a faint, almost pure, tone is audible for 1 up to 6 seconds. It is decaying and has a
sharp pitch in the spectral gap where no stimulus was available (Zwicker, 1964; Lummis and Guttman, 1972). Both the localization of the Zwicker tone in the brain
and its origin have been long-standing open problems. In terms of our model we would expect the cause of this auditory illusion to be the SR controlled upregulated internal noise. Even though, thresholds are not increased within the channel usually processing frequencies that are within the spectral gap of the presented noise, it is plausible that cross-talk between adjacent frequency channels plays a role here. The sound intensities, or auditory nerve firing rates respectively, of neighbouring channels may serve as some kind of reference. Another finding which is perfectly consistent with our model is the fact that during the Zwicker tone sensation, auditory sensitivity for tone pulses at frequencies adjacent to the Zwicker tone are improved by up to 13 dB (Wiegrebe et al.,  1996).

\section*{References}
Berlin CI, Hood L, Morlet T, Rose K, Brashears S (2003) Auditory neuropathy/dys-synchrony: Diagnosis and management. Mental Retardation and Developmental Disabilities Research Reviews 9:225-231.

Chen GD, Fechter LD (2003) The relationship between noise-induced hearing loss and hair cell loss in rats. Hear Res 177:81-90.

Coles RR (1984) Epidemiology of tinnitus: (1) prevalence. J Laryngol Otol Suppl 9:7-15.

Devarajan K, Gassner D, Durham D, Staecker H (2012) Effect of Noise Exposure Duration and Intensity on the Development of Tinnitus. In: ARO, p 593. Abs.

Eggermont JJ (2003) Central tinnitus. Auris Nasus Larynx 30 Suppl:S7-12.

Eggermont JJ (2013) Hearing loss, hyperacusis, or tinnitus: what is modeled in animal research? Hear Res 295:140-149.

Eggermont JJ, Roberts LE (2004) The neuroscience of tinnitus. Trends Neurosci 27:676-682.

Engineer ND, Riley JR, Seale JD, Vrana WA, Shetake JA, Sudanagunta SP, Borland MS, Kilgard MP (2011) Reversing pathological neural activity using targeted plasticity. Nature 470:101-104.

Engstrom B (1983) Stereocilia of sensory cells in normal and hearing impaired ears. A morphological, physiological and behavioural study. Scand Audiol Suppl 19:1-34.

Gerken GM (1996) Central tinnitus and lateral inhibition: an auditory brainstem model. Hear Res 97:75-83.

Heller AJ (2003) Classification and epidemiology of tinnitus. Otolaryngol Clin North Am 36:239-248.

Jastreboff PJ, Brennan JF, Sasaki CT (1988a) An animal model for tinnitus. Laryngoscope 98:280-286.

Jastreboff PJ, Brennan JF, Coleman JK, Sasaki CT (1988b) Phantom auditory sensation in rats: an animal model for tinnitus. Behav Neurosci 102:811-822.

Kirk CE, Smith DW (2003) Protection from acoustic trauma is not a primary function of the medial olivocochlear efferent system. J Assoc Res Otolaryngol 4:445-465.

Knipper M, Ruettiger L, Schick B, Dlugaiczyk J (2011) Glycine receptor agonists for the treatment of phantom phenomena. In: Google Patents.

Langguth B, Landgrebe M, Kleinjung T, Sand GP, Hajak G (2011) Tinnitus and depression. World J Biol Psychiatry 12:489-500.

Lewis JE, Stephens SD, McKenna L (1994) Tinnitus and suicide. Clin Otolaryngol Allied Sci 19:50-54.

Liberman LD, Suzuki J, Liberman MC (2015) Dynamics of cochlear synaptopathy after acoustic overexposure. J Assoc Res Otolaryngol 16:205-219.

Liberman MC, Kiang NY (1984) Single-neuron labeling and chronic cochlear pathology. IV. Stereocilia damage and alterations in rate- and phase-level functions. Hear Res 16:75-90.

Nowotny M, Remus M, Kossl M, Gaese BH (2011) Characterization of the perceived sound of trauma-induced tinnitus in gerbils. J Acoust Soc Am 130:2827-2834.

Ries DT (2007). The influence of noise type and level upon stochastic resonance in human audition. Hearing Research 228, 136-143

Ruttiger L, Singer W, Panford-Walsh R, Matsumoto M, Lee SC, Zuccotti A, Zimmermann U, Jaumann M, Rohbock K, Xiong H, Knipper M (2013) The reduced cochlear output and the failure to adapt the central auditory response causes tinnitus in noise exposed rats. PLoS One 8:e57247.

Schaette R, McAlpine D (2011) Tinnitus with a normal audiogram: physiological evidence for hidden hearing loss and computational model. J Neurosci 31:13452-13457.

Schaette R, Kempter R (2006) Development of tinnitus-related neuronal hyperactivity through homeostatic plasticity after hearing loss: a computational model. European Journal of Neuroscience, Vol. 23, 3124-3138

Turner JG (2007) Behavioral measures of tinnitus in laboratory animals. Prog Brain Res 166:147-156.

Turner JG, Brozoski TJ, Bauer CA, Parrish JL, Myers K, Hughes LF, Caspary DM (2006) Gap detection deficits in rats with tinnitus: a potential novel screening tool. Behav Neurosci 120:188-195.

Wang J, Powers NL, Hofstetter P, Trautwein P, Ding D, Salvi R (1997) Effects of selective inner hair cell loss on auditory nerve fiber threshold, tuning and spontaneous and driven discharge rate. Hearing Research 107:67-82.

Wiegrebe L, Koessl M, Schmidt S (1996) Auditory enhancement at the absolute threshold of hearing and its relationship to the Zwicker tone. Hearing Research 100, 171-180.

Zeng FG, Fu QJ, Morse R (2000). Human hearing enhanced by noise. Brain Research 869, 251-255.

Zheng XY, Henderson D, McFadden SL, Hu BH (1997a) The role of the cochlear efferent system in acquired resistance to noise-induced hearing loss. Hear Res 104:191-203.

Zheng XY, Henderson D, Hu BH, Ding DL, McFadden SL (1997b) The influence of the cochlear efferent system on chronic acoustic trauma. Hear Res 107:147-159.

Roberto Benzi, Alfonso Sutera, and Angelo Vulpiani. The mechanism of stochastic resonance. Journal of Physics A: mathematical and general, 14(11):L453, 1981.

Anthony N Burkitt. A review of the integrate-and-fire neuron model: I. homogeneous synaptic input.Biological cybernetics, 95(1):1{19, 2006.

JAMES J Collins, THOMAS T Imho, and PETER Grigg. Noise-enhanced information transmission in rat sa1 cutaneous mechanoreceptors via aperiodic stochastic resonance. Journal of Neurophysiology, 76(1):642-645, 1996.

JJ Collins, Carson C Chow, Thomas T Imho, et al. Stochastic resonance without tuning. Nature, 376(6537):236-238, 1995.

John K Douglass, Lon Wilkens, Eleni Pantazelou, Frank Moss, et al. Noise enhancement of information transfer in craysh mechanoreceptors by stochastic resonance. Nature, 365(6444):337{340, 1993.

A Aldo Faisal, Luc PJ Selen, and Daniel M Wolpert. Noise in the nervous system. Nature Reviews Neuroscience, 9(4):292{303, 2008.

Luca Gammaitoni, Peter Huanggi, Peter Jung, and Fabio Marchesoni. Stochastic resonance. Reviews of modern physics, 70(1):223, 1998.

L Lapique. Recherches quantitatives sur l'excitation electrique des nerfs traitee comme une polarization. J Physiol Pathol Gen, 9:620{635, 1907.

Jacob E Levin and John P Miller. Broadband neural encoding in the cricket cereal sensory system enhanced by stochastic resonance. Nature, 380(6570):165{168, 1996.

JCR Licklider. A duplex theory of pitch perception. The Journal of the Acoustical Society of America, 23(1):147{147, 1951.

Hiroyuki Mino. The effects of spontaneous random activity on information transmission in an auditory	brainstem neuron model. Entropy, 16(12):6654-6666, 2014.

Sanya Mitaim and Bart Kosko. Adaptive stochastic resonance. Proceedings of the IEEE, 86(11):2152{2183, 1998.

Sanya Mitaim and Bart Kosko. Adaptive stochastic resonance in noisy neurons based on mutual information. Neural Networks, IEEE Transactions on, 15(6):1526{1540, 2004.

Frank Moss, Lawrence M Ward, and Walter G Sannita. Stochastic resonance and sensory information processing: a tutorial and review of application. Clinical Neurophysiology, 115(2):267{281, 2004.

Claude Elwood Shannon and Warren Weaver. The mathematical theory of communication. University of Illinois Press, 1959.

GWenning and K Obermayer. Activity driven adaptive stochastic resonance. Physical review letters, 90(12):120602, 2003.

Kurt Wiesenfeld, Frank Moss, et al. Stochastic resonance and the benefits of noise: from ice ages to crayfish and squids. Nature, 373(6509):33{36, 1995.

Ting Zhou, Frank Moss, and Peter Jung. Escapetime distributions of a periodically modulated bistable system with noise. Physical Review A, 		42(6):3161, 1990.

E. Zwicker, J. Acoust. Soc. Am. 36, 2413?2415 (1964).

R.C. Lummis and N. Guttman, J. Acoust. Soc. Am. 51, 1930?1944 (1972).

%\bibliographystyle{plain}

%\bibliographystyle{unsrt}
%\bibliography{refs}

\end{document}